\documentclass{scrartcl}
\usepackage{algorithm}
\usepackage{amsmath,amssymb}
\usepackage{listings}
\usepackage{xcolor}

\usepackage{tikz}
\usetikzlibrary{shapes,arrows}
\usetikzlibrary{calc,fit}

\usepackage{graphicx}
\usepackage{subfigure}
\usepackage{hyperref}

\usepackage{authblk}

\usepackage{booktabs}
\usepackage[style=ieee,backend=bibtex,url=false,doi=false,isbn=false,date=year,maxbibnames=10,minbibnames=10,maxcitenames=10,mincitenames=10]{biblatex}
\bibliography{paper}
\definecolor{mygray}{RGB}{220,220,220} 
\definecolor{mygreen}{RGB}{0,100,0} 
\definecolor{mylilas}{RGB}{170,55,241}
\providecommand{\keywords}[1]{\textbf{\textit{Index terms---}} #1}

\begin{document}

% \title{ALAP---A numerical implementation of ALADIN with examples}

\title{ALADIN-$\alpha$ -- An open-source MATLAB toolbox for distributed non-convex optimization}

\author[1]{Alexander Engelmann*}

\author[2]{Yuning Jiang}
\author[1]{Henrieke Benner}
\author[1]{Ruchuan Ou}
\author[2]{Boris Houska}
\author[1,3]{Timm Faulwasser}

%\author[3]{Author Three}

\affil[1]{{Institute for Automation and Applied Informatics}, \protect\\
	 {Karlsruhe Institute of Technology}, {Karlsruhe}, {Germany}}

%\presentaddress{Timm Faulwasser is now with the Institute of Energy Systems, Energy Efficiency and Energy Economics, TU Dortmund University, Dortmund, Germany}

\affil[2]{{School of Information Science and Technology}, \protect\\ {ShanghaiTech University}, {{Shanghai}, {China}}}

\affil[3]{currently with the Institute of Energy Systems, Energy Efficiency and Energy Economics,  \protect\\ TU Dortmund University, Dortmund, Germany}
%
%\address[3]{Institute of Energy Systems, Energy Efficiency and Energy Economics, TU Dortmund University, Dortmund, Germany}

\maketitle

\begin{abstract}
This paper introduces an open-source software for \emph{distributed} and \emph{decentralized} non-convex optimization named ALADIN-$\alpha$. 
ALADIN-$\alpha$ is a MATLAB implementation of tailored variants of the Augmented
Lagrangian Alternating Direction Inexact Newton (ALADIN) algorithm. Its
user interface is convenient for rapid prototyping of non-convex distributed
optimization algorithms.
An improved version of the recently proposed bi-level variant of ALADIN is included  enabling \emph{decentralized} non-convex optimization with reduced information exchange.
A collection of examples from different applications fields including chemical engineering, robotics, and	power systems underpins the  potential of ALADIN-$\alpha$.
\end{abstract}

\keywords{Distributed Optimization, Decentralized Optimization,  Nonconvex Optimization, ALADIN, ADMM, Optimal Power Flow, Distributed Model Predictive Control}

%\jnlcitation{\cname{%
%\author{A. Engelmann}, 
%\author{Y. Jiang}, 
%\author{H. Benner}, 
%\author{R. Ou}, 
%\author{B. Houska}, and 
%\author{T. Faulwasser}} (\cyear{2020}), 
%\ctitle{ALADIN-$\alpha$ -- An open-source MATLAB toolbox for distributed non-convex optimization}, \cjournal{Opti}, \cvol{2017;00:1--6}.}

\footnotetext{\textbf{Abbreviations:} {ALADIN, Augmented Lagrangian Alternating Direction Inexact Newton; ADMM, Alternating Direction Method of Multipliers; QP, Quadratic Program; NLP, Nonlinear Program; LICQ, Linear Independence Constraint Qualification; OCP, Optimal Control Problem; CG, Conjugate Gradients}}

\section{Introduction}\label{sec1}
Distributed \emph{non-convex} optimization is of significant interest in various engineering domains. 
These domains range from electrical power systems,\cite{Engelmann2020c,Erseghe2014,Kim2000,DallAnese2013} transportation problems,\cite{Jiang2017} via machine learning,\cite{Boyd2011} to distributed control,\cite{Tippett2013,Jiang2017,Christofides2013,Stewart2010} and distributed estimation.\cite{Shi2010,Du2019,Rabbat2004,Kar2012} 
However, only few software toolboxes for distributed optimization are currently available.
Moreover, these toolboxes are typically  tailored to specific applications and often focus on \emph{convex} problems.
Examples comprise implementations of the Alternating Direction of Multipliers Method (ADMM)  from Boyd et al.;\cite{Boyd2011}$^,$\footnote{\url{https://web.stanford.edu/~boyd/papers/admm/}} an implementation of ADMM for consensus problems;\footnote{\url{http://users.isr.ist.utl.pt/~jmota/DADMM/}}
and a  tailored implementation of ADMM for Optimal Power Flow (OPF) problems in Guo et al.\cite{Guo2017}$^,$\footnote{\url{https://github.com/guojunyao419/OPF-ADMM}} 
However, there is a lack of multi-purpose software tools for distributed optimization  and, to the best of the authors' knowledge, there are no generic toolboxes for both distributed and decentralized non-convex optimization.

Notice that we distinguish \emph{parallel} and \emph{distributed} optimization. 
In \emph{parallel} optimization, the main motivations are computational speed-up or  computational tractability, while {reducing} the amount of communication and the amount of central coordination is typically of secondary importance (due to shared memory architectures). 
{In \emph{distributed} optimization,} the main goal is to minimize central coordination  and communication  (distributed memory architectures).
\emph{Decentralized} optimization additionally requires communication purely on a neighbor-to-neighbor basis.
This is especially relevant in multi-agent settings, where individual entities cooperate to the end of optimization, control, or estimation---e.g. in the context of cyber-physical systems, IoT, or embedded control.
\emph{Essentially decentralized} optimization softens the requirement of pure neighbor-to-neighbor communication by allowing the global summation of scalars.\cite{Engelmann2021}

For \emph{parallel} optimization efficient structure-exploiting tools exist. 
Classical tools include the \texttt{GALAHAD} software collection and in particular the \texttt{LANCELOT} algorithm, which is based on augmented Lagrangians and efficiently solves problems on shared-memory architectures.\cite{Gould2003} 
A closed-source parallel interior point software is \texttt{OOPS}.\cite{Gondzio2007}
The open-source package \texttt{qpDUNES} is tailored towards the time-wise decomposition of Quadratic Programs (QPs) arising in model predictive control.\cite{Frasch2015}
%For general QPs, the partially parallelizable solver \texttt{OSQP} seems promising.\cite{Stellato2020}
\texttt{PIPS} is a collection of algorithms solving structured linear programs, QPs, and general Nonlinear Programming Problems (NLPs) in parallel.\cite{Chiang2014,Lubin2011}
The software \texttt{HiOp} is tailored towards structured and very large-scale NLPs with few nonlinear constraints.
It is  based on interior point methods.\cite{Petra2019,Petra2019a}
Moreover, combining parallel linear algebra routines (e.g. \texttt{PARDISO})\cite{Schenk2001} with standard nonlinear programming solvers (e.g. \texttt{IPOPT})\cite{Wachter2006} also leads to partially parallel algorithms.\cite{Curtis2012,Kourounis2018}
The tools mentioned above are implemented in low-level languages such as \texttt{C} or \texttt{C++} leading to a high computational performance.
On the other hand, their focus is mainly computational speedup via parallel computing rather than distributed and decentralized optimization in a multi-agent setting.

Classical distributed and decentralized optimization algorithms based on Lagrangian relaxation such as dual decomposition or ADMM are guaranteed to converge only for very specific non-convexities typically appearing in the objective function of the optimization problems commonly at a sublinear/linear rate.\cite{Wang2019,Hong2016,He2012}
{In many multi-agent applications, however, the non-convexities occur in the constraints. This implies} that classical algorithms are not guaranteed to converge.\cite{Erseghe2014,Christofides2013}
One of the few algorithms exhibiting fast convergence guarantees in the non-convex case is the Augmented Lagrangian Alternating Direction Inexact Newton (ALADIN) algorithm.\cite{Houska2016}
Yet---up to now---a publicly available software implementation of ALADIN is missing. 

The present paper introduces an open-source MATLAB implementation of different ALADIN variants in the toolbox ALADIN-$\alpha$. 
It is intended for rapid prototyping and aims at user-friendliness. 
The only user-provided information are objective and constraint functions---derivatives and numerical solvers are generated automatically using algorithmic differentiation routines and external state-of-the-art NLP solvers.
A rich set of examples covering  problems from robotics, power systems, sensor networks and chemical engineering underpins the application potential of ALADIN-$\alpha$.
Besides the vanilla ALADIN algorithm, ALADIN-$\alpha$ covers recent  extensions including:
\begin{itemize}
	\item improved\footnote{The version of bi-level ALADIN given here is improved in the sense that we use improved versions of d-CG and d-ADMM from Engelmann et al..\cite{Engelmann2021} In contrast to a previous version,\cite{Engelmann2020c} these two algorithms rely on a unified sparsity framework and do not require a precomputation phase lowering communication demand. } decentralization of   bi-level ALADIN   with essentially decentralized, respectively, decentralized  variants of the Conjugate Gradient  method (d-CG) and the Alternating Direction of Multipliers Method (d-ADMM) as inner algorithms;\cite{Engelmann2020c,Engelmann2021}
%	\item BFGS Hessian approximation offering a means to reduce communication and computation;\cite{Engelmann2019}
	\item the nullspace ALADIN variant   reducing communication and coordination;\cite{Engelmann2020c} 
	\item a parametric implementation enabling distributed Model Predictive Control (MPC), and,
	\item  heuristics for Hessian regularization and parameter tuning for improving performance.
\end{itemize}
Moreover, we provide an implementation of ADMM based on the formulation of Houska et al.,\cite{Houska2016} which uses the same interface as ALADIN. 
This way, comparisons between ALADIN and ADMM are fostered.
Moreover, ALADIN-$\alpha$ can be executed {in parallel mode} via the MATLAB parallel computing toolbox.
This often leads to a substantial speed-up, for example, in distributed estimation problems.
A documentation and many application examples of ALADIN-$\alpha$ are available under \url{https://alexe15.github.io/ALADIN.m/}.
We remark that ALADIN-$\alpha$ intends to be a \emph{rapid prototyping} environment to enable  testing of distributed and decentralized algorithms for non-convex optimization based on ALADIN.
At this stage, computational speed or real-time feasibility are beyond the scope of the toolbox. 

%For achieving a high computation speed, other tools might be more appropriate.

The remainder of the  paper is organized as follows: \autoref{sec:probForm} recalls  the main ideas of ALADIN and bi-level ALADIN.
In \autoref{sec:toolbox} we comment on the code structure and data structures and present a simple tutorial example.
{Numerical examples  from chemical engineering, power systems, and sensor networks illustrate how to use ALADIN-$\alpha$ in different application domains in \autoref{sec:ex}.
The appendix provides implementation details.}

\section{Preliminaries} \label{sec:probForm}
We start  with a problem formulation amenable for distributed and decentralized optimization.

\subsection{Problem Formulation}
The ALADIN-$\alpha$ toolbox solves structured optimization problems of the form
\begin{subequations} \label{eq:sepProb}
	\begin{align} 
	\min_{x_1,\dots,x_{n_s}}\quad  & \sum_{i\in \mathcal{S}} f_i(x_i,p_i) \label{eq:SepProbObj} \\
	\;\;\text{subject to} \qquad &g_{i}(x_i,p_i) = 0 && \mid \kappa_i,  &\forall i \in \mathcal{S}, \label{eq:SepProbEqcnstr}\\
	&h_{i}(x_i,p_i) \leq 0  && \mid \gamma_i, & \forall i \in \mathcal{S}, \label{eq:SepProbInqcnstr}\\
	&\underline{x}_i \leq x_i \leq  \overline{x}_i&& \mid\eta_i,  &\forall i \in \mathcal{S},\label{eq:SepProbBoxCnstr} \\
	&\sum_{i\in \mathcal{S}}A_i x_i=b&& \mid\lambda,\label{eq:consCnstr}
	\end{align}
\end{subequations}
where $\mathcal{S}=\{1,\dots,n_s\}$ is a set of subproblems, $f_i:\mathbb{R}^{n_{xi}} \times \mathbb{R}^{n_{pi}}\rightarrow\mathbb{R}$ are objective functions, $g_i:\mathbb{R}^{n_{xi}} \times \mathbb{R}^{n_{pi}}\rightarrow\mathbb{R}^{n_{gi}}$ and  $h_i:\mathbb{R}^{n_{xi}} \times \mathbb{R}^{n_{pi}}\rightarrow\mathbb{R}^{n_{hi}}$ are  constraint functions of the subproblems $i \in \mathcal{S}$.
Upper and lower bounds $\underline x_i, \bar x_i \in \mathbb R^{n_{xi}}$ are considered separately for numerical efficiency reasons.
The matrices $A_i \in \mathbb{R}^{n_{c}\times n_{xi}}$ combined with $b \in \mathbb R^{n_c}$ model affine coupling constraints between the subproblems. The
Lagrange multipliers $\kappa$ assigned to the constraint $g$ are denoted by $g(x)=0  \quad|\; \kappa$.
The partially separable formulation of \eqref{eq:sepProb} is  generic: it contains several problem formulations as special cases such as consensus or sharing problems.
	Most NLPs can be reformulated in form of \eqref{eq:sepProb}  by introducing auxiliary variables.\cite{Boyd2011}
We discuss a particular reformulation example in \autoref{sec:ex}.
Note that problem~\eqref{eq:sepProb} allows for parametric problem data captured in  $p_i\in \mathbb{R}^{n_{pi}}$.
This  can be  useful in  MPC or if one would like to solve the same problem for varying  parameters.

\subsection{Standard and bi-level ALADIN} \label{sec:ALADIN}
\begin{figure}[h]
	\centering
	\includegraphics[width=\linewidth]{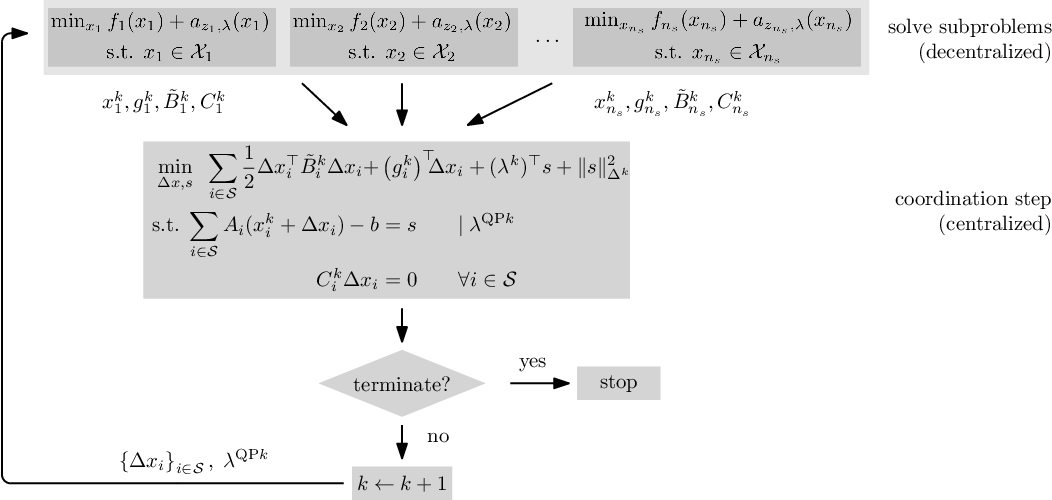}
	\caption{Simplified flow chart of standard ALADIN.}
	\label{fig:flowchart}
\end{figure}

ALADIN solves convex and non-convex optimization problems~\eqref{eq:sepProb} in a distributed fashion.
A simplified flow chart of standard ALADIN is sketched in \autoref{fig:flowchart}.
ALADIN combines ideas from ADMM and  Sequential Quadratic Programming (SQP) combining distributed computation from ADMM with fast convergence properties and guarantees from SQP.\cite{Houska2016}
Similar to ADMM, ALADIN adopts a parallel step---i.e., several NLPs are solved locally and in parallel to minimize local objective functions $f_i$ together with augmentation terms
\[
a_{z_i,\lambda}(x_i) = \lambda^\top A_i x_i + \left\|x_i-z_i\right\|_{\Sigma_i^k}^2.
\]
These terms account for the coupling between the subproblems.
	Here, $\Sigma_i \in \mathbb R^{n_{xi} \times n_{xi}} \succ 0$ are scaling matrices and $z_i \in \mathbb R^{n_{xi}}$ encodes the influence of other subproblems.
Moreover, local non-convex constraints
\[
\mathcal{X}_i = \{x_i \in  \mathbb{R}^{n_{xi}}\;|\; g_i(x_i)=0,\; h_i(x_i) \leq 0\}
\] are considered in each subproblem $i \in \mathcal{S}$.
Since these subproblem-specific NLPs are  similar in ALADIN and ADMM, both algorithms share the same computational complexity in the local step.
Sensitivities such as the gradients of the local objective $\nabla f_i(x_i)$, Hessian approximations $\tilde B_i$ and Jacobian matrices $(\nabla g_i,\nabla h_i)$ of the local constraints are evaluated locally.
These sensitivities are combined in a sparse coordination QP adopted from~SQP methods.
Note that the coordination QP is equality-constrained and strongly convex (under certain regularity assumptions)---thus it can be reformulated as a system of linear  equations.
The primal and dual solution vectors of this coordination QP are broadcasted to the local subproblems and the next ALADIN iteration starts.
The algorithm terminates once the norm of the violation of  the  constraint  \eqref{eq:consCnstr}  and the stepsize are both sufficiently small.

\begin{figure}
	\centering
	\includegraphics[width=\linewidth]{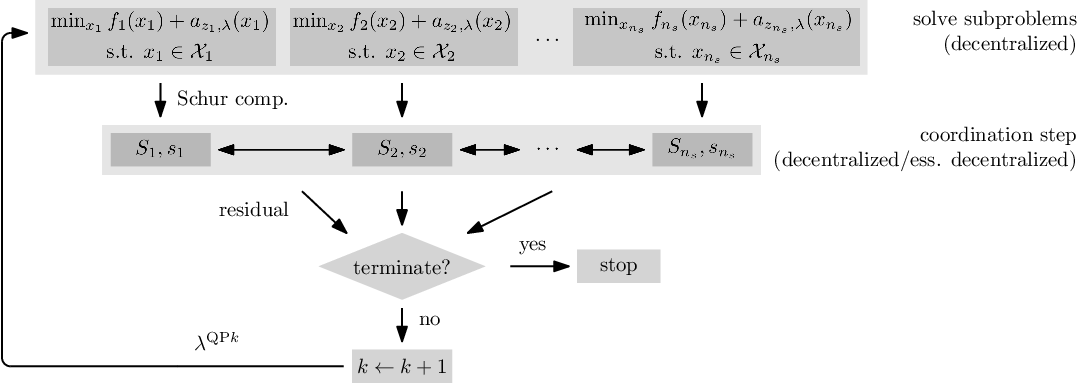}
	\caption{Simplified flow chart of bi-level ALADIN.}
	\label{fig:flowchartBil}
\end{figure}

The main advantage of standard ALADIN over other existing approaches are convergence guarantees and fast local convergence.\cite{Houska2016}
On the other hand, the coordination QP makes ALADIN distributed but \emph{not} decentralized.
Furthermore, the coordination step in standard ALADIN is quite heavy and communication intense compared with other algorithms such as ADMM.
Bi-level ALADIN overcomes these drawbacks by constructing a coordination QP of smaller dimension lowering communication.\cite{Engelmann2020c}
Here, the sensitivities are ``condensed'' by computing the Schur-complement of the KKT systems leading to $\{S_i\}, \{s_i\}$, which are of dimension $n_c$---the number of the coupling variables.
 The number of coupling variables is typically much smaller than the total number of variables.
These Schur-complements are combined in a lower-dimensional QP, which is solved in a decentralized fashion purely based on neighborhood communication leading to an overall decentralized algorithm.
A simplified flow chart of bi-level ALADIN is shown in \autoref{fig:flowchartBil}.
Observe that---in contrast to standard ALADIN (\autoref{fig:flowchart})---bi-level ALADIN solves the coordination QP in a decentralized fashion based decentralized inner algorithms.
ALADIN-$\alpha$ comes with two of these inner algorithms: an essentially decentralized version of the Conjugate Gradient~(d-CG) method and a decentralized version of ADMM~(d-ADMM).\cite{Engelmann2021}
The variables, which have to be exchanged in the solution process of the lower-dimensional QP depend on the particular decentralized algorithm at hand. 
Although these decentralized inner algorithms do not solve the coordination problem exactly, bi-level ALADIN is still guaranteed to converge locally under certain bounds on the numerical precision.\cite{Engelmann2020c}
A detailed description of ALADIN is given in Appendix~\ref{sec:ALADINdet}. 
%For the sake of completeness, more detailed descriptions of ALADIN, bi-level ALADIN, d-CG and d-ADMM are presented in  Appendix~\ref{sec:ALADINdet}.

% =================================================
% Set up a few colours
\colorlet{lcfree}{black}
\colorlet{lcnorm}{black}
\colorlet{lccong}{black}
% -------------------------------------------------
% Set up a new layer for the debugging marks, and make sure it is on
% top
\pgfdeclarelayer{marx}
\pgfsetlayers{main,marx}
% A macro for marking coordinates (specific to the coordinate naming
% scheme used here). Swap the following 2 definitions to deactivate
% marks.
\providecommand{\cmark}[2][]{%
	\begin{pgfonlayer}{marx}
		\node [nmark] at (c#2#1) {#2};
	\end{pgfonlayer}{marx}
} 
\providecommand{\cmark}[2][]{\relax}

\section{The ALADIN-$\alpha$ toolbox} \label{sec:toolbox}

This section presents the main contribution of this paper: the ALADIN-$\alpha$ toolbox implementing different ALADIN variants. 
We comment on its code structure and data structures.
Moreover, we illustrate the usage of ALADIN-$\alpha$ on a tutorial example.

\subsection{Code Structure} \label{sec:codeStruct}
\begin{figure}
	\centering
	\includegraphics[width=0.8\linewidth]{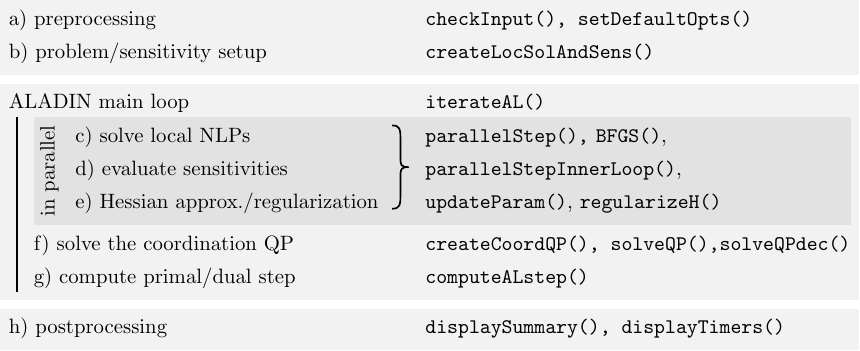}
	\caption{Structure  of \texttt{run\_ALADIN()} in ALADIN-$\alpha$.}
	\label{fig:codestruct}
\end{figure}

In order to simplify algorithm development and testing we choose a procedual/functional programming style.
All core features are implemented in MATLAB enabling easy rapid-prototyping.
The overall structure of \texttt{run\_ALADIN()}---the main function of ALADIN-$\alpha$---is shown in \autoref{fig:codestruct}.
First, a reprocessing step performs a consistency check of the input data and provides default options.
The \texttt{createLocSolAndSens()} function initializes the local NLPs and sensitivities for all subproblems $i \in \mathcal{S}$.
We use \texttt{CasADi}\cite{Andersson2019} for algorithmic differentiation and as an interface to many state-of-the-art NLP solvers such as  \texttt{IPOPT}.\cite{Wachter2006}
\texttt{CasADi} itself relies on pre-compiled code making function and derivative evaluation  fast. 
A \texttt{reuse} option avoids the reconstruction of the  \texttt{CasADi} problem setup, which enables the use of saved problem formulations.
When the \texttt{reuse} mode is activated (e.g. when ALADIN-$\alpha$ is used within an MPC loop), \texttt{createLocSolAndSens()}  is skipped, which results in a speed-up especially for large problems.

In the main loop \texttt{iterateAL()}, the function \texttt{parallelStep()} solves the local NLPs and evaluates the Hessian of the Lagrangian (or its approximation e.g. when BFGS is used), the gradient of the objective, and the Jacobian of the active constraints (sensitivities) at the NLP's solution.
The set of active constraints is determined by primal active set detection described in Appendix~\ref{sec:ALADINdet}.
Furthermore, a regularization procedure is executed if needed.
Moreover, in case the nullspace method or bi-level ALADIN is used, the computation of a nullspace basis and the computation of the Schur-complement is  performed locally shifting substantial computational burden from the centralized coordination step to \texttt{parallelStep()}.
The function \texttt{updateParam()} computes dynamically changing ALADIN parameters for numerical stability and speedup.

The coordination QP is constructed in the function \texttt{createCoordQP()}.
Different QP formulations are possible:
here we use a variant considering slack variables from Houska et al. for numerical stability.\cite{Houska2016} 
%Another option is to put the slacks on the constraint linearization $C^k_i \Delta x_i = 0$, or on both, or to entirely remove them.
Different dense and sparse solvers for solving the coordination QP are available  in \texttt{solveQP()}. 
Most of them are based on solving the first-order necessary conditions which is a system of linear equations. 
Available solvers are the MATLAB linear algebra routines \texttt{linsolve(), pinv()} and \texttt{MA57}.\footnote{MA57 is interfaced indirectly---we employ the MATLAB LDL factorization, which is based on MA57.}
Using sparse solvers can speed up the computation time substantially. 
Note that only \texttt{MA57} supports sparse matrices.
The solver can be specified by setting the \texttt{solveQP} option.
In case of convergence problems from remote starting points it can help to reduce the primal-dual stepsize of the QP step by setting the \texttt{stepSize} in the options to a value smaller than $1$.
More advanced step-size selection rules are subject to ongoing and future work.

\subsection{Data Structures} \label{sec:dataStr}
The  data structure for defining problems in form of \eqref{eq:sepProb} is a struct called \texttt{sProb}. 
This data structure collects the objective functions $\{f_i\}_{i \in \mathcal{S}}$ and constraint functions $\{g_i\}_{i \in \mathcal{S}}$ and $\{h_i\}_{i \in \mathcal{S}}$  in cells, which are contained in a nested struct called \texttt{locFuns}.
Furthermore, \texttt{sProb} collects lower/upper bounds  \eqref{eq:SepProbBoxCnstr} in cells  \texttt{llbx} and \texttt{uubx}.
The coupling matrices  $\{A_i\}_{i \in \mathcal{S}}$ are summarized in \texttt{AA}.
One can  provide NLP solvers and sensitivities optionally---in this case the problem construction in  \texttt{createLocSolAndSens()} is skipped leading to a  speedup  for large problems.
This way, problem setups can be saved and reused.
For a minimal working example of ALADIN-$\alpha$, one only needs to specify \texttt{ffi} and $\texttt{AA}$.
Optionally one can provide initial guesses \texttt{zz0} and initial Lagrange multipliers \texttt{lam0}.
The second ingredient for ALADIN-$\alpha$ is an \texttt{opts} struct, which specifies the ALADIN variant  and algorithm parameters. A full list of options with descriptions can be found in the code documentation.\footnote{\url{https://alexe15.github.io/ALADIN.m/options/}}

\begin{figure}
	\centering
	\includegraphics[width=0.7\linewidth]{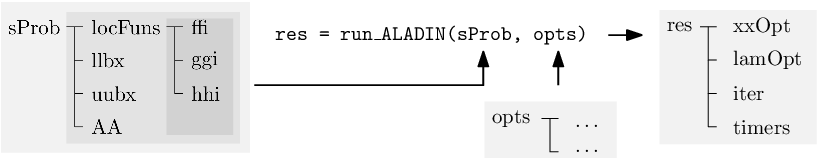}
	\caption{The \texttt{sProb} data structure for defining problems in form of \eqref{eq:sepProb}.}
	\label{fig:sprob}
\end{figure}

ALADIN-$\alpha$ returns a struct as output.
This struct contains the cell \texttt{xxOpt} with local minimizers $\{x_i^\star\}_{i \in \mathcal{S}}$ and the optimal Lagrange multipliers  $\lambda^\star$ of the consensus constraints \eqref{eq:consCnstr}.
Moreover the field \texttt{iter} contains information about the ALADIN iterates such as primal/dual iterates and \texttt{timers} collects timing information. 
Note that  \texttt{run\_ALADIN()} and \texttt{run\_ADMM()} have the same function signature in terms of \texttt{sProb}---only the options differ.

\subsection{Further Features}

We describe selected features of ALADIN-$\alpha$---a full description of all features can be found under \footnote{\url{https://alexe15.github.io/ALADIN.m/options/}}.
\\

\noindent
\textbf{Hessian Approximations}\;
Instead of exact Hessians, approximations such as the Broyden-Fletcher–Goldfarb-Shanno-(BFGS) update can be used either to reduce communication and/or to reduce computational complexity in sensitivity computation.
The BFGS Hessian is activated by setting the \texttt{Hess} option either to \texttt{BFGS} for standard BFGS or to \texttt{DBFGS} for damped BFGS. For details on BFGS we refer to the book of Nocedal and Wright.\cite{Nocedal2006}
\\

\noindent 
\textbf{Parametric NLP Setup}\;
A parametric problem setup, where the objective functions $f_i$ and the equality/inequality constraints $g_i/h_i$ depend on parameters $p_i$ is possible.
This feature is  useful in combination with the \texttt{reuse} option which returns the internally constructed \texttt{CasADi} solvers and derivatives.
If one provides a previously constructed NLP as input argument when calling  \texttt{run\_ALADIN()}, the problem construction is skipped, which can lead to a substantial speedup.
In an  MPC setting, for example, the parameter $p_i$ models the changing initial condition in the MPC loop.
Moreover, parametric  problem data might  be useful for large-scale problems where one would like to solve an optimization problem for a wide range of parameters.
This feature is activated by adding a parameter cell \texttt{p} to  \texttt{sProb} and defining the objective/constraints in terms of two inputs, $x_i$ and $p_i$.
An example illustrating how to use these features for distributed predictive control of two mobile robots is given in in the code repository.\footnote{\url{https://alexe15.github.io/ALADIN.m/robotEx/}}
\\

\noindent
\textbf{Parallelization}\; 
ALADIN-$\alpha$ also supports parallel computing on multiple processors via the MATLAB parallel computing toolbox.
Here, we exploit the fact that the local NLPs are independent from each other, i.e., they can be solved in parallel.
An example for distributed nonlinear estimation with mobile sensor networks can be found in \autoref{sec:fusion}.
Parallel computing can be activated by setting the \texttt{parfor} option to \texttt{true}.
\\
\begin{table}[b]
	\centering
	\begin{tabular}{rllll}
		\toprule
		example & field  & \texttt{examples/...} & docs \\
		\midrule 
		chemical reactors & chemical engineering/control & \texttt{chemical\_reactor} \\
		mobile robots &  robotics/control&\texttt{robots} & \footnotesize {\url{https://alexe15.github.io/ALADIN.m/robotEx/}}\\
		optimal power flow & power systems& \texttt{optimal\_power\_flow}& \footnotesize{\url{https://alexe15.github.io/ALADIN.m/redComm/}}\\
		sensor network & estimation  & \texttt{sensor\_network}& \footnotesize{\url{https://alexe15.github.io/ALADIN.m/ParallelExample/}} \\
		\bottomrule 
	\end{tabular}
	\caption{Application examples coming with ALADIN-$\alpha$.	}
	\label{tab:Apps}
\end{table}

\noindent
\textbf{Application Examples}\; 
We provide numerical examples highlighting applicability of ALADIN-$\alpha$ to a wide range of problems.
The code for all these examples is available in the \texttt{examples\textbackslash } folder of ALADIN-$\alpha$.
Furthermore, we provide  descriptions of these examples in the documentation online.\footnote{\url{https://alexe15.github.io/ALADIN.m/}}
Beyond the  examples of this section, we consider distributed optimal control  and the application of ALADIN-$\alpha$ to test problems from the Hock-Schittkowski test collection in the online repository.\cite{Engelmann2020c,Hock1980,Mehrez2017}
A list of all examples is given in \autoref{tab:Apps}.

\subsection{A Tutorial Example}
Consider the non-convex NLP
\begin{subequations} \label{eq:tutProb}
	\begin{align} 
	\min_{x_1,x_2\in \mathbb{R}} & \;f(x)=2 \, (x_1 - 1)^2 + (x_2 - 2)^2\label{eq:TutProbObj} \\
	\text{subject to}&\quad -1 \leq x_1 \cdot x_2 \leq 1.5. \label{eq:TutProbInEqcnstr}
	\end{align}
\end{subequations}
In order to apply ALADIN-$\alpha$, we reformulate problem  \eqref{eq:tutProb} in form of \eqref{eq:sepProb}.
We introduce auxiliary variables $y_1, y_2$ with $y_1 \in \mathbb{R}$ and  $y_2= (y_{21}\; \; y_{22})^\top $.
We couple these variables again by introducing a  consensus constraint $\sum_i A_i y_i = 0$ with $ A_1 = 1 $ and $A_2= (-1\;\; 0)$.
Furthermore, we reformulate the objective function $f$ by local objective functions $ f_1(y_1) : = 2\,  (y_1 - 1)^2$ and $f_2(y_2) =  (y_{22} - 2)^2$ with $f=f_1 + f_2$.
Moreover, reformulate the global inequality constraint \eqref{eq:TutProbInEqcnstr} by a local two dimensional constraint  $h_2 = (h_{21}\; h_{22})^\top$ with $h_{21}(y_2) = -1 - y_{21}\, y_{22}$ and $h_{22}(y_2) =  -1.5 + y_{21}\, y_{22}$.
Combining these reformulations yields
\begin{subequations}  \label{eq:qProbRef}
	\begin{align}  
		\min_{y_1 \in\mathbb{R},y_2 \in \mathbb{R}^2}   2 \,(y_{1} - 1)^2 &+   (y_{22} - 2)^2\\
	\;\;\text{subject to} \;\;      -1 - y_{21}\,y_{22} &\leq 0, \quad 
	-1.5 + y_{21} y_{22} \leq 0, \\
	y_1 + (\,-1 \;\; 0 \,)\,y_2 &= 0,
	\end{align}
\end{subequations}
which is in form of \eqref{eq:sepProb}.
Note that the solutions to  \eqref{eq:tutProb}  and \eqref{eq:qProbRef} coincide but \eqref{eq:qProbRef} is of higher dimension.
This reformulation reveals a general strategy for reformulating problems in form of \eqref{eq:sepProb}: if there is  nonlinear coupling in the objective functions or the constraints, introduce auxiliary variables and require them to coincide by an additional consensus constraint in form of \eqref{eq:consCnstr}.
\\~\\

\noindent
\textbf{Solution with ALADIN-$\alpha$}
Next, we transcribe \eqref{eq:qProbRef} in the struct \texttt{sProb} as illustrated in \autoref{sec:dataStr}.
To highlight different possibilities of problem setup, we construct the problem in three different ways: a) via the MATLAB symbolic toolbox, b) via the \texttt{CasADi} symbolic framework and, c) directly via function handles, cf. \autoref{cd:exCode}.

\begin{figure}
	\begin{center}
		\begin{minipage}[t]{.28\textwidth} 
			\begin{lstlisting}[language=Matlab, numbers=none, backgroundcolor = \color{mygray},linewidth=4.8cm]
% define symbolic variables
y1  =   sym('y1',[1,1],'real');
y2  =   sym('y2',[2,1],'real');

% define symbolic objectives
f1s =   2*(y1-1)^2;
f2s =   (y2(2)-2)^2;

% define symbolic ineq. constraints

h2s = [  -1-y2(1)*y2(2); ...
-1.5+y2(1)*y2(2)];

% convert symbolic variables to MATLAB fuctions
f1 = matlabFunction(f1s,'Vars',{y1});
f2 = matlabFunction(f2s,'Vars',{y2});

h1 = @(y1)[];
h2 = matlabFunction(h2s,'Vars',{y2});
			\end{lstlisting}
		\end{minipage} \hspace{.1cm}
		\begin{minipage}[t]{.28\textwidth} 
			\begin{lstlisting}[language=Matlab, numbers=none, backgroundcolor = \color{mygray},linewidth=4.5cm]
% define symbolic variables
y_1 = SX.sym('y_1', 1);
y_2 = SX.sym('y_2', 2);

% define symbolic objectives
f1s = 2 * (y_1 - 1)^2;
f2s = (y_2(2) - 2)^2;

% define symbolic ineq. constraints
h1s = [];
h2s = [  -1 - y_2(1)*y_2(2); ...
-1.5 + y_2(1)*y_2(2)]; 

% convert symbolic variables to MATLAB fuctions
f1 = Function('f1', {y_1}, {f1s});
f2 = Function('f2', {y_2}, {f2s});

h1 = Function('h1', {y_1}, {h1s});
h2 = Function('h2', {y_2}, {h2s});
			\end{lstlisting}
		\end{minipage}\hspace{.2cm}
		\begin{minipage}[t]{.28\textwidth} 
			\begin{lstlisting}[language=Matlab, numbers=none, backgroundcolor = \color{mygray},linewidth=4.5cm]




% define objectives
f1 = @(y1) 2 * (y1 - 1)^2;
f2 = @(y2) (y2(2) - 2)^2;

% define inequality constraints
h1 = @(y1) []
h2 = @(y2) [-1 - y2(1) * y2(2);...
-1.5 + y2(1) * y2(2)];








			\end{lstlisting}
		\end{minipage}
	\end{center}	
	\caption{Problem setup via MATLAB symbolic (left), CasADi (middle), and function handles (right).}\label{cd:exCode}
\end{figure}

\begin{figure}[h]
	\centering
	\includegraphics[width=\linewidth]{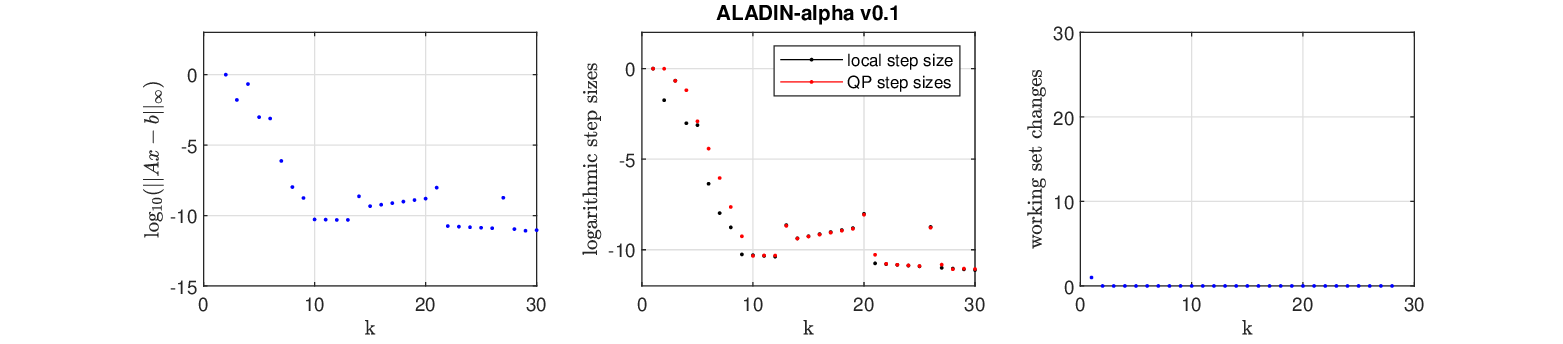}
	\caption{ALADIN-$\alpha$ iteration plot for tutorial problem \eqref{eq:qProbRef}. }
	\label{fig:microexout}
\end{figure}

After defining objective and constraint functions, all function handles and the coupling matrices $A_i$ are collected in \texttt{sProb}.
We call \texttt{run\_ALADIN()} with an empty options struct leading to computation with default parameters.
The code and the resulting ALADIN-$\alpha$ report after running \texttt{run\_ALADIN()} are shown in \autoref{cd:output}.
In the ALADIN-$\alpha$ report, the reason for termination and timing information is displayed. 
\autoref{fig:microexout} shows the output of ALADIN-$\alpha$ while it is running.
The figures show (in this order) the consensus violation $\|Ax-b\|_\infty$, the local step sizes $\|x^k-z^k\|_\infty$, the step size in the coordination step $\|\Delta x^k\|_\infty$, and  the changes in the active set.
Note that online plotting may consume a substantial amount of time---hence it is advisable to deactivate online plotting if there is not required e.g. for diagnostic reasons.

\begin{figure}
	\begin{center}
		\begin{minipage}[t]{.2\textwidth} 
			\begin{lstlisting}[language=Matlab, numbers=none, backgroundcolor = \color{mygray},linewidth=6cm]
% define coupling matrices
A1 = 1;
A2 = [-1, 0];

% collect problem data in sProb struct
sProb.locFuns.ffi = {f1, f2};
sProb.locFuns.hhi = {h1, h2};

% handing over of coupling matrices to problem
sProb.AA = {A1, A2};

% start solver with default options
sol = run_ALADINnew(sProb);











			\end{lstlisting}
		\end{minipage} \hspace{.2cm}
		\begin{minipage}[t]{.4\textwidth} 
			\begin{lstlisting}[language=Matlab, numbers=none, backgroundcolor = \color{mygray},linewidth=7.5cm]
========================================================      
==              This is ALADIN-alpha v0.1             ==      
========================================================      
QP solver:        MA57
Local solver:     ipopt
Inner algorithm:  none

No termination criterion was specified.
Consensus violation: 6.6531e-12

Maximum number of iterations reached.

---------------   ALADIN-alpha timing   ----------------
t[s]                %tot       %iter                
Tot time......:     3.92                                                         
Prob setup....:     0.19                4.8                                      
Iter time.....:     3.72                95                                        
---------                                                                       
NLP time......:     1.1                             29.7                 
QP time.......:     0.11                             2.8                 
Reg time......:     0.02                             0.6                 
Plot time.....:     2.27                            60.8                 

========================================================
			\end{lstlisting} 	
		\end{minipage}
	\end{center}	
	\caption{Collection of variables (left) and output of ALADIN-$\alpha$ (right).}\label{cd:output}
\end{figure}

\section{Numerical case studies} \label{sec:ex}

We present three case studies to shed light on  the differences of the implemented algorithms.
We consider an optimal control problem for a chemical reactor, an OPF problem, and a sensor localization example.

\subsection{Distributed Optimal Control of a Chemical Process System}
\begin{figure}[t]
	\centering
	\includegraphics[width=200pt]{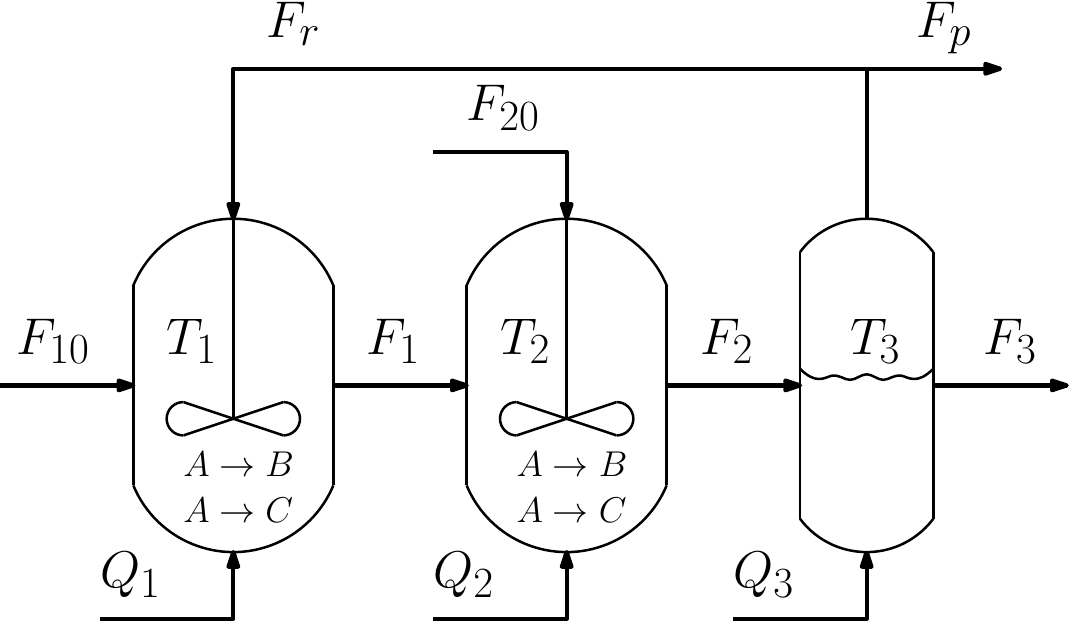}
	\caption{Reactor-separator process. }
	\label{fig:process}
\end{figure}

We consider a discrete-time optimal control problem (OCP) for a chemical process system.
This OCP can serve as a basis for distributed model predictive control.\cite{Rawlings2019,Stewart2011,Muller2017}
The process  consists of two Continuous Stirred-Tank Reactors (CSTRs) and a flash separator  shown in \autoref{fig:process}.\cite{Cai2014,Christofides2011} 
The goal is to steer the system to the optimal setpoint
\begin{align*}
u_s^\top  = 
\begin{pmatrix}
0 & 0 & 0
\end{pmatrix}\;
\text{ and } \;
x_s^\top  = 
\begin{pmatrix}
369.53& 3.31& 0.17& 0.04& 435.25& 2.75& 0.45& 0.11&  \\
435.25& 2.88 & 0.50& 0.12 
\end{pmatrix}
\end{align*}
from  $x(0)^\top  = \left (360.69\;\; 3.19\;\; 0.15\;\; 0.03\;\; 430.91\;\; 2.76\;\; 0.34\;\; 0.08\;\; 430.42\;\; 2.79\;\; 0.38\;\; 0.08 \right )$.
After applying a fourth-order Runge-Kutta scheme for  discretization, the dynamics of all CSTRs and the flash separator  are given by 
\begin{align*}
\label{eq:dynamic}
x_i^{k+1} = q_i(x_i^k,u^k_i,z_i^k) \;\; \text{ for } \;\;  i\in \mathcal{S} ,
\end{align*}
where $q_i: \mathbb{R}^{n_{xi}} \times \mathbb{R}^{n_{ui}} \times \mathbb{R}^{n_{zi}} \rightarrow \mathbb{R}^{n_{xi}}$ are the dynamics of the $i$th vessel with $\mathcal{S} :=\{1,2,3\}$ being the set of vessels.
Here, $x_i^\top =(x_{Ai},x_{Bi},x_{Ci},T_i)$ are the states, $x_{Ai},x_{Bi},x_{Ci} $  are the concentrations of the reactants $A$, $B$ and $C$, and $T$ is the temperature.
The input $u_i=Q_i$ denotes the heat-influx of the $i$th vessel and $z_i:= (x_j)_{j \in N(i)}$ are copied states of all neighbors $N(i) \subseteq \mathcal{S}$.
Note that the feed-stream flow rates $F_{10},F_{20},F_3,F_R$ and $F_p$ are fixed and given.
A detailed description of the system dynamics is given in Christofides et al.\cite{Christofides2011}
With the above, we  formulate a discrete-time optimal control problem 
\begin{subequations} \label{eq:dOCP}
	\begin{align}  
	\min_{\substack{(x_i^k,\; z_i^k, u_i^k), k\in \mathbb{I}_{[1\;\;T]} \; \\  i\in \mathcal{S} }}  \sum_{i\in \mathcal{S}} \sum_{k\in \mathbb{I}_{[1\;\;T]}}\frac{1}{2}  (x_i^k-x_{is})^\top Q_i^{\phantom k} &(x_i^k -x_{is}) \;+ \; \frac{1}{2} (u_i^k-u_{is})^\top R_i^{\phantom k} (u_i^k - u_{is}^k)   \\
	\text{s.t.} \quad x_i^{k+1} - q_i(x_i^k,u_i^k,z_i^k) =0, \quad x_i^0 = x_i(0)\;\;  &\text{ for all } k \in \mathbb{I}_{[1\;\;T]}\; \text{ and for all } i \in \mathcal{S},  \\
	\underline u_i \leq u_i^k \leq \bar u_i, \quad  	\underline x_i \leq x_i^k,  \quad    &\text{ for all } k \in \mathbb{I}_{[1\;\;T]}\; \text{ and for all } i \in \mathcal{S}, \\
	\sum_{i\in \mathcal{S}}A_i \left (x_i^{k\top}  \; z_i^{k\top } \;  u_i^{k\top} \right )^\top=0 \quad &\text{ for all } k \in \mathbb{I}_{[1\;\;T]},
	\end{align}
\end{subequations}
with lower/upper bounds on the inputs $\overline{u} = -\underline{u} = (5\cdot   10^4\;\; 1.5 \cdot  10^5\;\; 2 \cdot  10^5)^\top  $, and lower bounds on the states $\underline{x}_i^k = 0$ for all times $k \in \mathbb{I}_{[1\;\;T]}$ and all vessels $i \in \mathcal{S}$.
The weighting matrices are $Q_i = \text{diag}(20\;\;10^3\;\;10^3\;\;10^3)$ and $R_i = 10^{-10}$.
The matrices $A_i$ are constructed to model the constraint $z_i:= (x_j)_{j \in N(i)}$. 
The sampling time is $\Delta h=0.01h$ and the horizon is $T=10\,$h.
By defining $\tilde x_i^\top :=\left (x_i^{k\top}  \; z_i^{k\top } \;  u_i^{k\top} \right )_{k \in \mathbb{I}_{[1\;\;T]}}$, 
$f_i(\tilde x_i):=\sum_{k\in \mathbb{I}_{[1\;\;T]}} \frac{1}{2}  (x_i^k-x_{is})^\top Q_i^{\phantom k} (x_i^k -x_{is}) + \sum_{k\in \mathbb{I}_{[1\;\;T-1]}} \frac{1}{2} (u_i^k-u_{is})^\top R_i^{\phantom k} (u_i^k - u_{is}^k)$,
$g_i(\tilde x_i):=  \left (x_i^{k+1} - q_i(x_i^k,u_i^k,z_i^k) \right )_{k \in \mathbb{I}_{[1\;\;T-1]}}$, and 
$h_i(\tilde x_i):= \left ( (\underline u_i - u_i^k \; \; u_i^k - \bar u_i\; \; \underline x_i - x_i^k)^\top \right )_{k \in \mathbb{I}_{[1\;\;T]}}$ the OCP \eqref{eq:dOCP} is in form of \eqref{eq:sepProb}, where $\tilde x_i$   corresponds to $x_i$ in \eqref{eq:sepProb}.\\~\\

\begin{figure}[h]
	\centering
	\subfigure[States]{
		\includegraphics[trim={2.5cm 0 1.3cm 0},clip,width=0.48\linewidth]{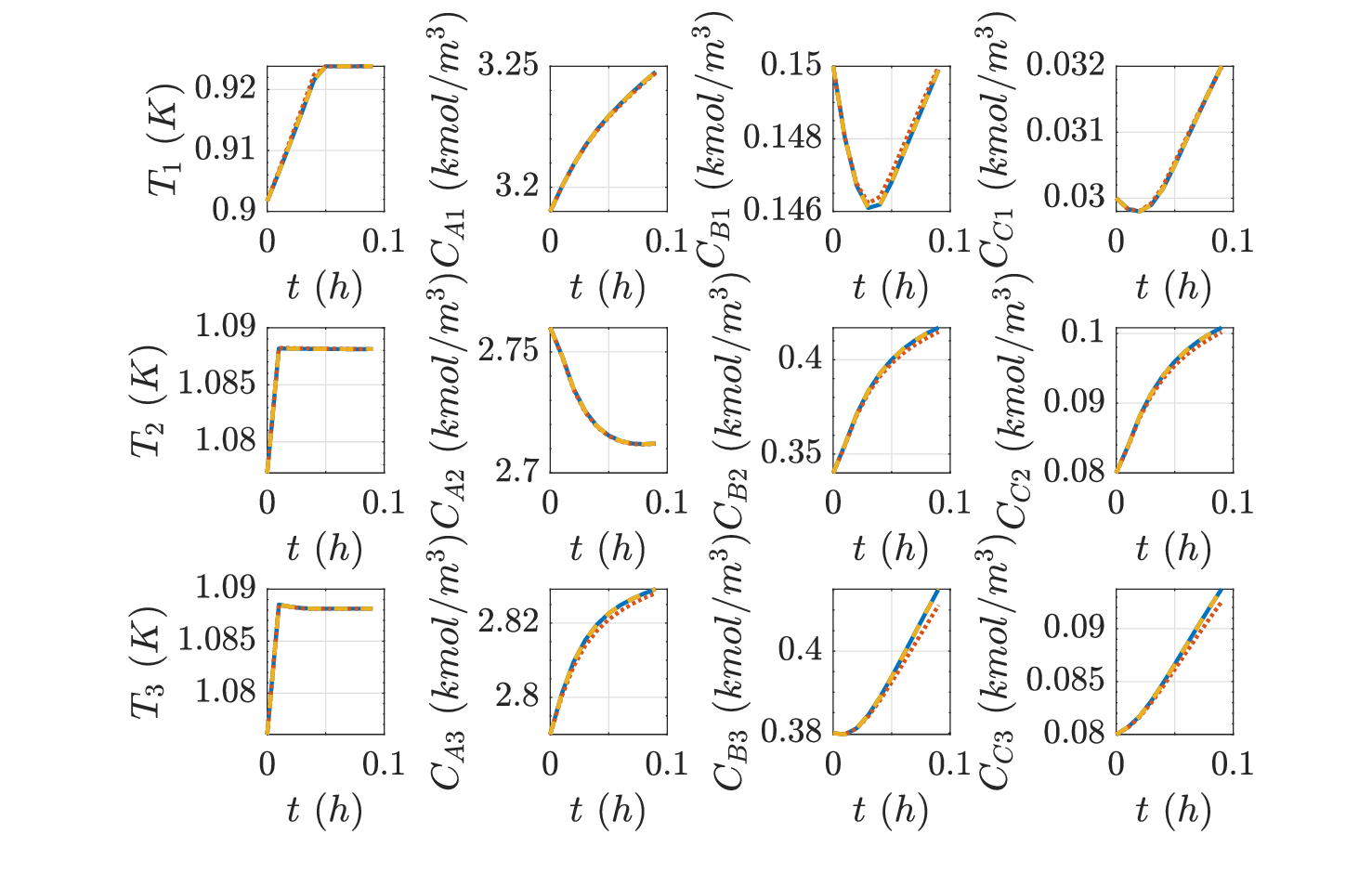}
	}
	\subfigure[Inputs]{
		\includegraphics[width=0.48\linewidth]{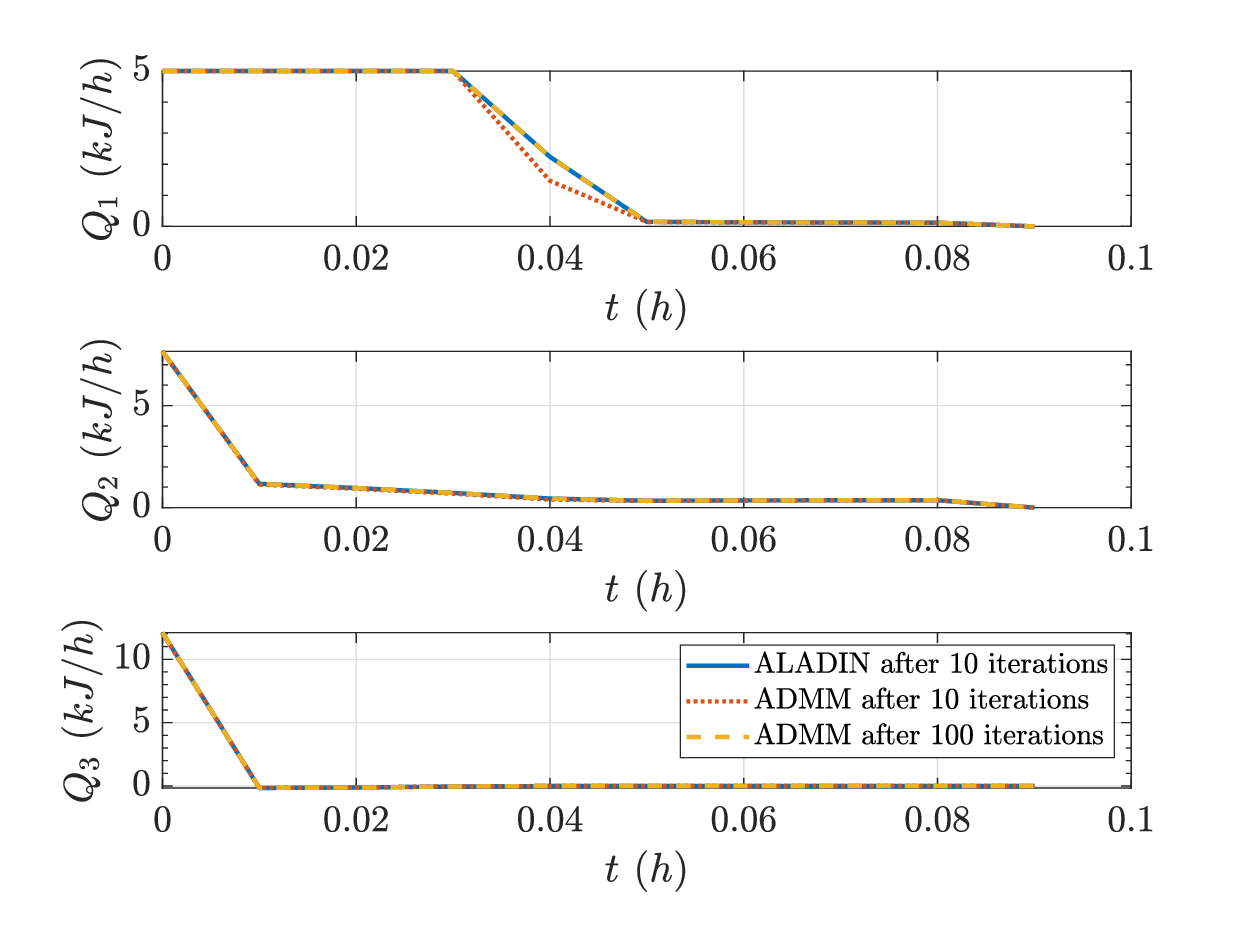}
	}
	\caption{Optimal state/input trajectories computed by ALADIN \& ADMM. }
	\label{fig:chem_aladin}
\end{figure}

%\begin{figure}
%	\centering
%	\includegraphics[width=.85\linewidth]{figures/chem_convergence.eps}
%	\caption{Convergence of ALADIN (solid lines) and ADMM (dashed lines) for problem \eqref{eq:dOCP}.}
%	\label{fig:convAL}
%\end{figure}

\begin{figure}[h]
	\centering
	\includegraphics[width=\linewidth]{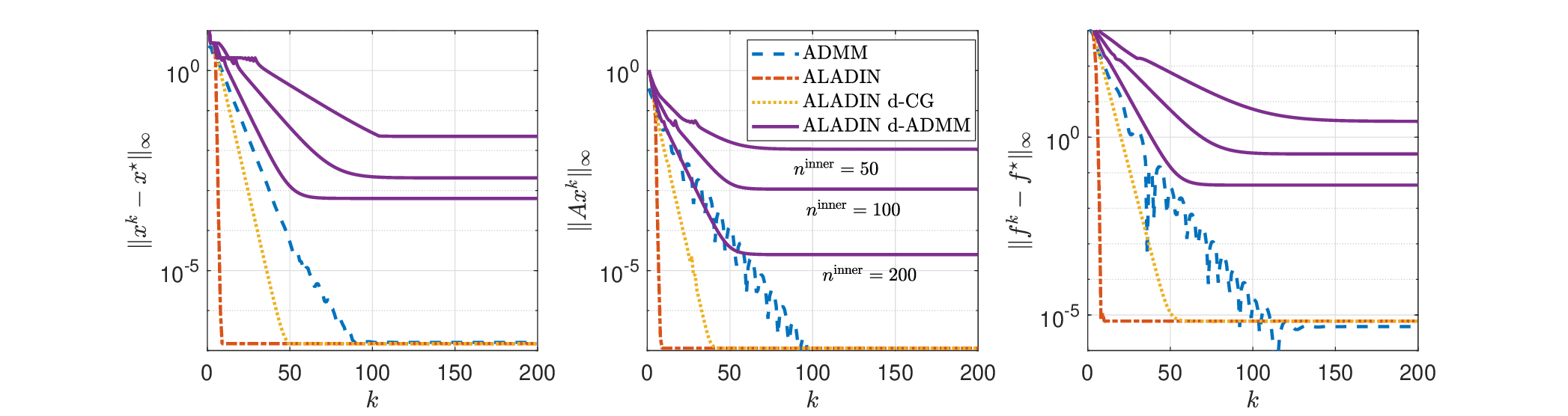}
	\caption{Numerical performance of ALADIN, bi-level ALADIN, and  ADMM for  OCP~\eqref{eq:dOCP}. }
	\label{fig:chem}
\end{figure}

\textbf{Numerical Results} 
\autoref{fig:chem} shows the convergence behavior of standard ALADIN, of bi-level ALADIN with decentralized conjugate gradients (d-CG) as inner algorithm, of bi-level ALADIN with decentralized ADMM (d-ADMM) as inner algorithm, and of  ADMM over the iteration index $k$.
Specifically, we depict the distance to a minimizer $\|x^k-x^\star\|_\infty$, the consensus violation $\|Ax^k-b\|_\infty$, and the optimality gap $|f(x^k)-f(x^\star)|$.
Note that for the considered problem, ADMM is not guaranteed to converge because of the nonlinear dynamics. 
However, since ADMM is nevertheless used in may works, we use it as a baseline for comparison.\cite{Bestler2019,Tang2019,Farokhi2014}
Bi-level ALADIN with d-ADMM is executed with  inner d-ADMM iterations $n^{\mathrm{inner}}\in \{50,100,200\}$ and bi-level ALADIN with d-CG is executed with $200$ inner d-CG iterations.
One can see that  ADMM converges   fast and there seems to be no benefit when using bi-level ALADIN with ADMM as an inner algorithm.
Basic ALADIN and ALADIN with conjugate gradients converges faster, but one has to solve an expensive coordination step in case of basic ALADIN or to perform many inner iterations in case of bi-level ALADIN with d-CG.

\autoref{fig:chem_aladin} shows the resulting open-loop  input and state trajectories for  OCP~\eqref{eq:dOCP} for ALADIN and ADMM after 20 iterations, and for ADMM after 100 iterations. 
At first-glance, all trajectories look quite similar.
However, small differences in the input trajectories can be observed. 
Close inspection of  \autoref{fig:chem} shows that in logarithmic scale the differences can be  large.
For example the consensus gap $\|Ax-b\|_\infty$ is in an order of $10^{-1}$ after 20 iterations, which means that the physical values at the interconnection points have a maximum mismatch of $10^{-1}$.

%
%}

\subsection{Distributed Optimal Power Flow}
Next, we consider an OPF problem, which is one of the most important optimization problems in power systems.\cite{Frank2016}
Distributed optimization is particularly important here due to large  problem sizes and due to the necessity of a reduced information exchange between subsystems. 
\begin{figure}[h]
	\centering
	\includegraphics[width=0.7\linewidth]{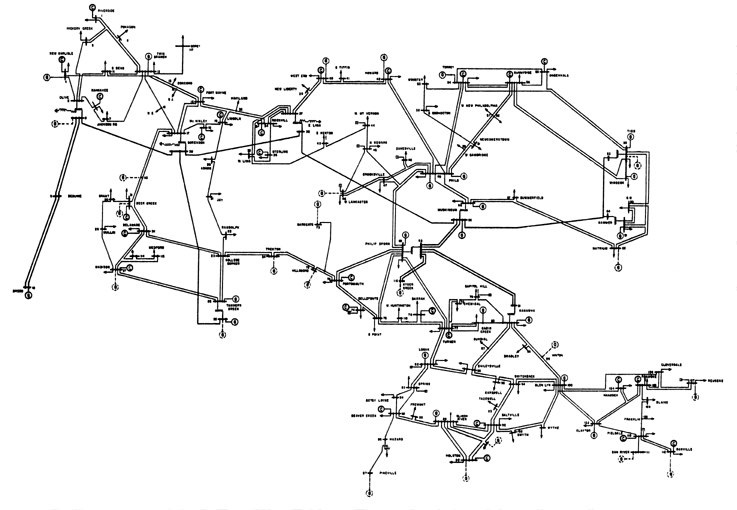}
	\caption{Map of the IEEE 118-bus test system.}
	\label{fig:118bus0}
\end{figure}
We consider the IEEE 118-bus test case shown in \autoref{fig:118bus0}, which comprises  about 500 decision variables.
A detailed problem description to match \eqref{eq:sepProb} is beyond the scope of this paper.  Details on  this and on the  partitioning scheme are given in Engelmann et al.\cite{Engelmann2019}

\begin{figure}
	\centering
	\includegraphics[width=\linewidth]{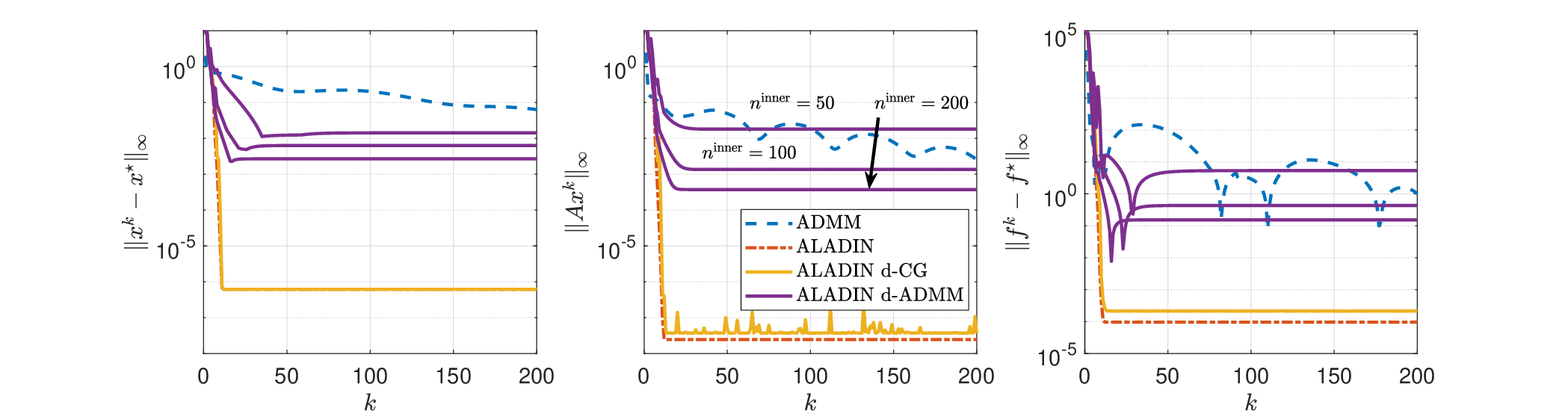}
	\caption{Numerical performance of ALADIN, bi-level ALADIN and  ADMM for the 118-bus OPF problem.}
	\label{fig:118bus}
\end{figure}

	\textbf{Numerical Results} 
\autoref{fig:118bus} shows the performance of all  distributed and decentralized optimization algorithms coming with ALADIN-$\alpha$.
Bi-level ALADIN with d-ADMM is executed with  inner d-ADMM iterations $n^{\mathrm{inner}}\in \{50,100,200\}$ and bi-level ALADIN with d-CG is executed with $70$ inner iterations.
One can see that in contrast to the chemical reactor from the previous subsection ADMM converges quite slowly and  requires about 1,500 iterations to converge to an acceptable level of accuracy.
This underlines that the performance of ADMM is  problem dependent---especially in a setting with non-convex constraints.
Basic ALADIN and bi-level ALADIN with d-CG on the other hand converge rapidly and to a high accuracy.
For bi-level ALADIN with d-ADMM, the achievable accuracy depends on the number of inner d-ADMM iterations.

\autoref{tab:timing} shows  timing information for all algorithms converging to $\|Ax^k\|_\infty < 10^{-3}$.
We use a computer with an Intel Core i7-8550U processor with 4 cores, 16\,GiB of memory, and MATLAB R2020a running Arch Linux with parallel computing disabled.  
The initialization phase of the sensitivities is not considered. 
One can see that  there is not much difference between the  ALADIN variants since most of the time is spent in solving the local NLPs and this step is the same.
ADMM is about five times slower since it requires many more iterations and thus many more NLP solutions are computed.

\begin{table}[t]
	\centering
	\begin{tabular}{lccl}
		\toprule
		basic ALADIN & \multicolumn{2}{c}{bi-level ALADIN} &  ADMM \\
		& d-CG (70) & d-ADMM (200) & \\
				\midrule 
		$2,5$s & 2,8s & 4,5s & 16,2s \\
		\bottomrule 
	\end{tabular}
	\caption{Timings for different algorithms converging to  $\|Ax^k\|_\infty < 10^{-3}$ .	}
	\label{tab:timing}
\end{table}

\section{Summary \& Future Work}
This paper has introduced  one of the first open source toolboxes for distributed non-convex optimization: ALADIN-$\alpha$.
It is based on the Augmented Lagrangian Alternating Direction Inexact Newton (ALADIN) algorithm and implements various extensions mostly aiming at reducing communication and coordination overhead. 
Moreover, {{$\text{ALADIN-}\alpha$}} comes with a rich set of examples from different engineering fields reaching from power systems over non-linear control to mobile sensor networks.

Although ALADIN-$\alpha$ performs  well for many small to medium-sized problems, we aim at further improving numerical stability in future work by developing more advanced internal auto-tuning routines. 
Furthermore, developing distributed globalization strategies for enlarging the set of possible initializations seems important and promising.
Code generation for distributed optimization on embedded devices is another interesting research direction. 
A possible alternative to ALADIN-based schemes for distributed non-convex optimization seem essentially decentralized interior point methods.\cite{Engelmann2021,Engelmann2021a}

\textbf{Acknowledgement}
We would like to thank Tillmann M\"uhlpfordt for very helpful discussions and suggestions, and Veit~Hagenmeyer for  supporting the development of ALADIN-$\alpha$.
Moreover, Timm Faulwasser acknowledges financial support by the Elite Program for Postdocs of the Baden-Württemberg Stiftung. 
\appendix

\color{white}
\section{Implementation Details} \label{sec:Feat}
\color{black}

The main ALADIN algorithm is based on Houska et al.\cite{Houska2016}, but additional practical considerations have been taken into account improving efficiency and numerical stability.
\subsection{ALADIN in Detail} \label{sec:ALADINdet}
Standard ALADIN is summarized in Algorithm~\ref{alg:ALADIN}.
\begin{algorithm}[t]
	\caption{Augmented Lagrangian Alternating Direction Inexact Newton (ALADIN)}
	\textbf{Initialization:} Initial guess $\left (\{z_i^0\}_{i\in \mathcal{S}},\lambda^0 \right )$, choose $\{\Sigma_i^k\} \succ 0,\{\Delta^k\} \succ 0,\tau>0,\epsilon>0$. \\
	\textbf{Repeat:}
	\begin{enumerate}
		\item \textit{Parallelizable Step:} \label{step:parStep}
		For each $i \in \mathcal{S}$, solve  
		\begin{align}
		\label{eq::denlp}
		\underset{x_i}{\min}& f_i(x_i) + (\lambda^k)^\top A_i x_i + \left\|x_i-z_i^k\right\|_{\Sigma_i^k}^2 
		\text{s.t.}\; g_i(x_i) = 0,  h_i(x_i)\leq 0, \underline{x}_i \leq x_i \leq  \overline{x}_i.
		\end{align}
		\item \textit{Termination Criterion:} If 
		$
		\label{eq::stop}
		\left\|\sum_{i\in \mathcal{S}}A_ix^k_i -b \right\|\leq \epsilon \text{ and } \left\| x^k - z^k \right \|\leq \epsilon\;,
		$
		 $x^\star = x^k$.
		
		\item \textit{Sensitivity Evaluation:} Compute and communicate gradients,
		Hessian approximations  and constraint Jacobians according to \eqref{eq:HessApprox} and \eqref{eq:Jac}, respectively. 
		
		\item \textit{Consensus Step:} Solve the coordination QP 
		\begin{align} 
		\notag
		\underset{\Delta x,s}{\min}\;\;\sum_{i\in \mathcal{S}}\frac{1}{2}\Delta x_i^\top \tilde B^k_i\Delta x_i + &\left (g_i^k \right )^{\top} \Delta x_i     + (\lambda^k)^\top s + \|s\|^2_{\Delta^k}  \\ 
		\text{subject to}\;                                   \sum_{i\in \mathcal{S}}A_i(x^k_i+\Delta x_i)-b &=  s   \label{eq::conqp}  \qquad  |\; \lambda^{\mathrm{QP} k},\\
		C^k_i \Delta x_i &= 0                                     \qquad   \forall i\in \mathcal{S}, \notag
		\end{align}
		and return $\Delta x^k$ and $\lambda^{\mathrm{QP}k}$.

		\item \textit{Line Search:} Update primal and dual variables by
		\begin{eqnarray}\notag
		 	z^{k+1}\leftarrow x^k + \alpha^k \Delta x^k \qquad \qquad
		\lambda^{k+1}\leftarrow\lambda^k + \alpha^k (\lambda^{\mathrm{QP}k}-\lambda^k),
		\end{eqnarray}
		with $\alpha^k=1$ for a full-step variant. 
		Update $\Sigma_i^k$ and $\Delta^k$.   
	\end{enumerate}
	\label{alg:ALADIN}
\end{algorithm}
Each ALADIN iteration executes three main steps: step 1. solves local NLPs \eqref{eq::denlp} for fixed and given values for primal iterates $z_i^k$ and dual iterates $\lambda^k$ in parallel.
The parameter sequences $\{\Sigma_i^k\} \succ 0$ and $\{\Delta^k\} \succ 0$ are user-defined---details are described in\autoref{sec:scalMat}.\footnote{We use scaled 2-norms $\|x\|_\Sigma:=\sqrt{x^\top \Sigma x}$ for $\Sigma \succ 0$ here.}  
Note that the equality constraints~\eqref{eq:SepProbEqcnstr} and box constraints~\eqref{eq:SepProbBoxCnstr} are not explicitly detailed in Houska et al.\cite{Houska2016}---we consider them separately here for numerical efficiency reasons.\footnote{Some numerical solvers for \eqref{eq::denlp} can for example treat box constraints in an efficient way by using projection methods.}
Step 2. of Algorithm~\ref{alg:ALADIN} computes sensitivities such as the gradients of the  objective functions $g_i^k=\nabla f_i(x_i^k)$ and positive definite approximations of the local Hessian matrices 
\begin{equation}\label{eq:HessApprox}
\tilde B_i^k \approx B_i^k =  \nabla^2_{xx}\big \{f_i(x_i^k)+\kappa_i^{k\top } g_i(x_i) + \left (\gamma_i^{k\top} \;\; \eta_i^{k\top }\right ) _{j \in \mathcal{A}_i^k   } \left ( \tilde h_j(x_i^k) \right )_{j \in \mathcal{A}_i^k} \big \},
\end{equation}
where
\begin{align*}
\mathbb{A}_i^k :=\left \{ j \in \{1,\dots, n_{hi} + 2n_{xi}\} \; | \; \left (\tilde h(x^k) \right )_j > -\tau     \right    \}
\end{align*}
is the set of active inequality constraints  in subproblems $i \in \mathcal{S}$ and $\tau >0$ is a user-defined parameter which can be specified via the \texttt{actMargin} option.
Moreover, we define combined inequality constraints 
\begin{equation}
\tilde h(x)^\top := \left (h(x)^\top \;\; (\underline{x}_i -x_i )^\top \;\; ( x_i - \bar {x}_i)^\top  \right ),
\end{equation}
and Jacobians of  active constraints
\begin{equation} \label{eq:Jac}
C^{k\top}_i :=\big (\nabla g_i(x^k_i)^\top\;  \left (\nabla \tilde  h_j(x^k_i) \right )_{j\in \mathbb{A}_i^k}^\top \big )
\end{equation}
for all $i \in \mathcal{S}$.
With this  information, step 4. of Algorithm~\ref{alg:ALADIN} solves an equality constrained quadratic program \eqref{eq::conqp} serving as a coordination problem.
%\footnote{When $B_i^k$ is positive definite (which is guaranteed by Hessian modifications), \eqref{eq::conqp} is strongly convex and can be can be replaced by it's first-order optimality conditions.}
Step 5. of Algorithm~\ref{alg:ALADIN} updates $z^k$ and $\lambda^k$ based on the solution to  \eqref{eq::conqp}. 
To achieve  global convergence guarantees, the step size parameter $\alpha \in (0\;\;1]$ has to be properly chosen by a globalization routine.
Designing suitable distributed globalization routines is subject of ongoing and future work---we use the full step variant  $\alpha=1$.
A smaller stepsize can be specified via the \texttt{stepSize} option which might stabilize ALADIN-$\alpha$ for certain problems.
Note that time-varying parameter sequences $\{\Delta^k\}$ and $\{\Sigma_i^k\}$ with $\Sigma_i^k, \Delta^k \succ 0$ might accelerate convergence of ALADIN in practice.
Heuristic routines for doing so are described in \autoref{sec:scalMat}.\footnote{Note that in contrast to Houska et al.\cite{Houska2016}, we omit the term $\rho/2$ in front of the penalization term in~\eqref{eq::denlp} avoiding redundancy. The setting from Houska et al.\cite{Houska2016} can be recovered by choosing $\Sigma_i^k=\rho^k/2\,I$.}

\subsection{Solving the Coordination QP}
The Hessian approximations $B_i^k$ are assumed to be positive definite.
Hence, problem \eqref{eq::conqp} is a strictly convex equality-constrainted QP which can be solved via the first order optimality conditions (if $C_i^k$ has full row rank) which is a system of  linear equations.
There are two possibilities for solving \eqref{eq::conqp} numerically: either by centralized linear algebra routines or by iterative methods.
For centralized computation, several solvers are interfaced in ALADIN-$\alpha$ which can be specified via the \texttt{solveQP} option.
The available solvers are summarized in \autoref{tab:QPsol}.
Note that not all solvers support sparse matrices.
MA57 usually perfoms very well in practice---both in terms of speed and robustness. 
The second approach to solve \eqref{eq::conqp} is via iterative and \emph{decentralized} routines such as d-CG and d-ADMM.
Details of these decentralized routines are described in \autoref{sec:bilAL}.

 \begin{table}[b]
	\centering
	\begin{tabular}{rllll}
		\toprule
		name	& MA57  &      pinv  & linsolve&  \\ \midrule 	
		algorithm & multifrontal LDL  & based on SVD & LU  \\
		sparse & yes & yes & no \\
\bottomrule
	\end{tabular}
	\caption{Centralized QP solvers interfaced in ALADIN-$\alpha$.	}
	\label{tab:QPsol}
\end{table}

\subsection{Hessian Approximations} \label{sec:hessApp}
As $B_i^k$ may have zero eigenvalues  or may even be indefinite if evaluated via \eqref{eq:HessApprox}, special care has to be taken.\footnote{If one would use \eqref{eq:HessApprox} regardless, the coordination step \eqref{eq::conqp} would not necessarily produce descent directions  destroying the local convergence properties of ALADIN. In case of zero eigenvalues, $B_i^k$ is singular and the coordination step can not be solved by a standard solver for linear systems of equations.}
Here we use a heuristic using ideas from Nocedal and Wright\cite{Nocedal2006}---other heuristics are possible and might accelerate convergence.
Our heuristic ``flips'' the sign of the negative eigenvalues (if there are any) and puts the zero eigenvalues to a small positive number~$\delta$.
The intuition here is  that the stepsize in the direction of negative curvature becomes smaller the ``more negative'' the curvature is.
For doing so we compute the eigendecomposition $B_i^k=V_i \Lambda_i V_i^\top$ for each subproblem $i$ locally, where $\Lambda_i$ is a matrix with the eigenvalues of $H_i$ on its main diagonal and $V_i$ is the matrix eigenvectors. 
Hence, the regularization reads 
\begin{align*}
\tilde{\Lambda}_{jj} :=
\begin{cases}
|{\Lambda}_{jj}| & \text{if } {\Lambda}_{jj} < - \delta,\\
\delta & \text{if } |{\Lambda}_{jj}| \in (-\delta, \;\delta), \\
{\Lambda}_{jj} & \text{else},
\end{cases}
\qquad 
\text{and}
\qquad 
\tilde B_i^k:=V_i \tilde \Lambda_i V_i^\top,
\end{align*}
with $\delta = 10^{-4}$.
Regularization  can be activated by the option \texttt{reg} and $\delta$ can be specified via \texttt{regParam}.

As an alternative to exact Hessians with regularization, one can use the Broyden-Fletcher–Goldfarb-Shanno (BFGS) update for successively approximating the exact Hessian based on the gradient of the Lagrangian.
This has the advantage that only the gradient of the Lagrangian has to be communicated (which is a vector) instead of the Hessian (which is a matrix).
A detailed description on how to use BFGS within ALADIN can be found in Engelmann et al.\cite{Engelmann2019} 
The BFGS formula can be activated by the setting the option \texttt{Hess} to \texttt{BFGS} or to \texttt{DBFGS} for damped BFGS. 
The advantage of damped-BFGS is that it guarantees positive-definiteness of $B_i^k$ regardless of the positive-definiteness of the exact Hessian at the current iterate.
Note that in case the nullspace method is used (cf. \autoref{sec:redSp}), the regularization is done for the reduced Hessian $\bar B_i^k$ instead of $B_i^k$.
%Moreover, Gauss-Newton variants of ALADIN for least-squares-type optimization problems from estimation are possible \cite{Du2019}.

\subsection{Scaling Matrices} \label{sec:scalMat}
A simple heuristic for the sequences $\{\Sigma_i^k\} \succ 0$ and $\{\Delta^k\} \succ 0$ is to start with certain (usually diagonal) initial matrices $\Sigma_i^0, \Delta^0$ and to multiply them by a factor $r_\Delta,r_\Sigma > 1$ in each iteration, i.e. 
\begin{align} \label{eq:SigUpdate}
\Sigma_i^{k+1} =
\begin{cases}
r_\Sigma \Sigma_i^k\   &\text{if} \; \|\Sigma_i\|_\infty < \bar \sigma\\ 
\Sigma_i^k&\text{otherwise} 
\end{cases} 
\qquad   \text{and} \qquad  
\Delta^{k+1} =
\begin{cases}
r_\Delta \Delta ^k\   &\text{if} \; \|\Delta\|_\infty < \bar \delta\\ 
\Delta^k&\text{otherwise} .
\end{cases} 
\end{align}
These routines have been successfully used in  previous works.\cite{Engelmann2019,Engelmann2020c}
An alternative for choosing  $\{\Delta^k\}$ is based on the consensus violation for each individual row in \eqref{eq:consCnstr}.
The idea here is to increase the corresponding $\Delta _{ii}^k$ to drive the corresponding consensus violation to zero.
This technique is common in algorithms based on augmented Lagrangians, cf. Bertsekas\cite[Chap 4.2.2]{Bertsekas1999}.
Mathematically this means that we choose
\begin{align} \label{eq:rhoMuUpdate}
\Delta_{cc}^{k+1} =
\begin{cases}
\beta \, \Delta_{cc}^k   &\text{if} \; \left |\left (\sum_{i\in \mathcal{S}}A_i x_i^k - b \right )_c \right | >\gamma \left  |\left (\sum_{i\in \mathcal{S}}A_i x_i^{k-1} - b \right )_c \right | \phantom{\bigg |}\\
\Delta^k_{cc}&\text{if} \;  \left |\left (\sum_{i\in \mathcal{S}}A_i x_i^k - b \right )_c \right | \leq \gamma \left  |\left (\sum_{i\in \mathcal{S}}A_i x_i^{k-1} - b \right )_c \right | \phantom{\bigg |}
\end{cases} 
\; \forall \;
c\in 1,\dots,n_c,
\end{align}
with $\gamma \in (0\; \;1)$ and $\beta > 1$.
In ALADIN-$\alpha$ we choose $\beta= 10$ and $\gamma = 0.25$.
This rule can be activated by the option \texttt{DelUp} and is able to accelerate convergence of ALADIN-$\alpha$ substantially in some cases.

Note that the above heuristics such as regularization or parameter updates do not interfere with the fast local convergence properties of ALADIN-$\alpha$.
They are required for guaranteeing fast local convergence since they ensure that the assumptions made in the local convergence proof of ALADIN such as the positive-defniteness of $\tilde B_i^k$ are satisfied.\cite{Houska2016}

\subsection{The Nullspace Method} \label{sec:redSp}

The nullspace method can be activated to reduce the dimensionality of the coordination QP \eqref{eq::conqp}, thus reducing communication and computation in the coordination step.
The idea  is to parameterize the nullspace of the active constraints $\operatorname{null}(C_i^k):=\{x_i \in \mathbb{R}^{n_{xi}} \; |\; C_i^k x_i^k=0\}$ by $\operatorname{null}(C_i^k)= Z_i^k\Delta v_i$, where $Z_i^k \in \mathbb{R}^{x_{xi} \times\left  (n_{xi} - |\mathbb{A}_i^k|\right )}$ is matrix whose columns are a basis of  $\operatorname{null}(C_i^k)$.
Note that  $C_i^kZ_i^k=0$ by definition of the nullspace.
Using this parametrization, \eqref{eq::conqp} can be written as 
\begin{align}
\begin{aligned} 
&\underset{\Delta v,s}{\min}\;\;\sum_{i\in \mathcal{S}}\left\{\frac{1}{2}\Delta v_i^\top \bar B^k_i\Delta v_i + \bar  g_i^{k\top} \Delta v_i\right\}     + (\lambda^k)^\top s + \|s\|^2_{\Delta^k}  \\ 
&\label{eq:redQP} 
\text{subject to}\;                                   \sum_{i\in \mathcal{S}}\bar A^k_i(v^k_i+\Delta v_i) - b =  s    \qquad  |\; \lambda^{\mathrm{QP} k},
\end{aligned}
\end{align}
where $\bar B_i^k=Z_i^{k\top}B_i^k Z_i^k \in \mathbb{R}^{\left (n_{xi} - |\mathbb{A}_i^k| \right )\times \left (n_{xi} - |\mathbb{A}_i^k| \right )}$, $\bar g_i^{k} = Z_i^{k\top}g_i^k \in \mathbb{R}^{\left (n_{xi} - |\mathbb{A}_i^k| \right )}$ and $\bar A_i^k= A_i Z_i^k\in \mathbb{R}^{n_c \times \left (n_{xi} - |\mathbb{A}_i^k| \right )}$.
Note that $\bar A_i^k$ has an iteration index $k$ and changes during the iterations since $Z_i^k$ changes.
Similar to the full-space approach, regularization from \autoref{sec:hessApp} is used (if it is activated via the option  \texttt{reg}) yielding a positive definite $\bar B_i^k$.
The nullspace method can be used by activating the option \texttt{redSpace}.
Notice that  the required communication between the subproblems and the coordinator is reduced by twice the number of equality constraints and active inequality constraints.
Thus, the communication reduction can be  large for problems with many constraints.
Furthermore, the coordination QP \eqref{eq::conqp} is in general less expensive to solve since \eqref{eq:redQP} is of smaller dimension than \eqref{eq::conqp}.
Indeed,  \eqref{eq:redQP} is strongly convex under suitable assumptions which \eqref{eq::conqp} is not necessarily.\cite{Engelmann2020c} 
While computing nullspaces is numerically expensive (due to singular-value decomposition),  it is done parallel in our context---thus fostering parallelization.

\subsection{Bi-level ALADIN} \label{sec:bilAL}
Bi-level ALADIN is an extension of ALADIN to further reduce dimensionality of the   coordination QP \eqref{eq:redQP}. 
Moreover, it enables the use of decentralized ADMM or essentially decentralized conjugate gradients as  inner algorithms leading to an overall (essentially) \emph{decentralized} ALADIN variant.

We briefly recall the main idea of bi-level ALADIN.
Under the assumptions from Engelmann et al.,\cite{Engelmann2020c} evaluating the KKT conditions for \eqref{eq:redQP}  yields 
\begin{subequations} \label{eq:KKTred}
	\begin{align} 
	\bar B^k \Delta v + \bar g^k + \bar A ^{k\top} \lambda^{\mathrm{QP}}&=0, \label{eq:SchurElim} \\
	\bar A^k(v^k + \Delta v) -b - \frac{1}{2}\left (\Delta^k \right )^{-1}(\lambda^{\mathrm{QP}}-\lambda^k) &=0, \label{eq:redKKTdelV}
	\end{align}
\end{subequations}
where $\bar B^k$, $\bar A^k$ and $\bar \Delta v^k$ are block-diagonal concatenations of $\bar B_i^k$, $\bar A_i^k$ and $\bar \Delta v_i^k$.
Using the \emph{Schur-complement} reveals that \eqref{eq:KKTred} is equivalent to solving the system of linear equations
\begin{align} \label{eq:SchurComp}
\left  (\sum_{i \in \mathcal{S}} S_i^k+\; \frac{1}{2}\left (\Delta^k \right )^{-1}  \right  ) \lambda^{\mathrm{QP}}=\sum_{i \in \mathcal{S}} s_i^k   + \left (\frac{1}{2} \left (\Delta^k \right )^{-1} \lambda^k - b \right ) ,
\end{align}
where  $S_i^k := \bar A_i^k \, \bar B_i^{k^{-1}} \bar A_i^{k\top} \succ 0$ are local Schur-complement matrices and $s_i^k:= \bar A_i^k\left (v_i^k - \bar  B_i^{k^{-1}} \bar g_i^k\right )$ are local Schur-complement vectors.
The key observation for decentralization is that the matrices $S_i^k$ and vectors $s_i^k$ inherit the sparsity pattern of the consensus matrices $A_i$, i.e., zero rows in $A_i$ yield zero rows/columns in $S_i^k$ and $s_i^k$.
Intuitively speaking, each row/column of $S_i^k$ corresponds to one consensus constraint (row of \eqref{eq:consCnstr}) and only the subproblems which ``participate'' in this constraint have non-zero rows in their corresponding $S_i$.
This sparsity can be exploited to solve \eqref{eq:SchurComp} in a decentralized fashion.
 Examples for such algorithms are decentralized ADMM or an essentially decentralized  conjugate gradients algorithm  presented in Engelmann and Faulwasser.\cite{Engelmann2021}

\noindent

\printbibliography

%\bibliographystyle{natbib}
% the address of the bibtexfile might be wrong, so I modify it --Ruchuan Ou
%\bibliography{../bibtexfile/paper} % 
%\bibliography{zotero} % 

\end{document}